\documentclass[
  reprint,
  superscriptaddress,
  amsmath,amssymb,
  aps,
  prapplied,
]{revtex4-2}
\usepackage{graphicx}
\usepackage{dcolumn}
\usepackage{bm}
\usepackage{siunitx}
\usepackage{xcolor}
\usepackage{hyperref}
\usepackage{natbib}

\usepackage[normalem]{ulem}

\def\load{}
\def\znw{Z_\mathrm{nw}}
\DeclareMathOperator\arctanh{arctanh}

\DeclareSIUnit{\belmilliwatt}{Bm}
\DeclareSIUnit{\dBm}{\deci\belmilliwatt}
\begin{document}
\title{Rapid microwave-only characterization and readout of quantum dots using multiplexed gigahertz-frequency resonators}

\author{Damaz de Jong}
\affiliation{QuTech and Kavli Institute of Nanoscience, Delft University of Technology, 2600 GA Delft, The Netherlands}

\author{Christian Prosko}
\affiliation{QuTech and Kavli Institute of Nanoscience, Delft University of Technology, 2600 GA Delft, The Netherlands}

\author{Daan M. A. Waardenburg}
\affiliation{QuTech and Kavli Institute of Nanoscience, Delft University of Technology, 2600 GA Delft, The Netherlands}

\author{Lin Han}
\affiliation{QuTech and Kavli Institute of Nanoscience, Delft University of Technology, 2600 GA Delft, The Netherlands}

\author{Filip K. Malinowski}
\affiliation{QuTech and Kavli Institute of Nanoscience, Delft University of Technology, 2600 GA Delft, The Netherlands}

\author{Peter Krogstrup}
\affiliation{Center for Quantum Devices, Niels Bohr Institute, University of Copenhagen \& Microsoft Quantum Materials Lab Copenhagen, Denmark}

\author{Leo P. Kouwenhoven}
\affiliation{QuTech and Kavli Institute of Nanoscience, Delft University of Technology, 2600 GA Delft, The Netherlands}
\affiliation{Microsoft Quantum Lab Delft, Delft University of Technology, 2600 GA Delft, The Netherlands}
\author{Jonne V. Koski}
\affiliation{Microsoft Quantum Lab Delft, Delft University of Technology, 2600 GA Delft, The Netherlands}
\author{Wolfgang Pfaff}
\affiliation{Microsoft Quantum Lab Delft, Delft University of Technology, 2600 GA Delft, The Netherlands}
\affiliation{Department of Physics and Frederick Seitz Materials Research Laboratory, University of Illinois at Urbana-Champaign, Urbana, IL 61801, USA}

\date{\today}

\begin{abstract}
Superconducting resonators enable fast characterization and readout of mesoscopic quantum devices. 
Finding ways to perform measurements of interest on such devices using resonators only is therefore of great practical relevance.
We report the experimental investigation of an InAs nanowire multi-quantum dot device by probing GHz resonators connected to the device. 
First, we demonstrate accurate extraction of the DC conductance from measurements of the high-frequency admittance. 
Because our technique does not rely on DC calibration, it could potentially obviate the need for DC measurements in semiconductor qubit devices. 
Second, we demonstrate multiplexed gate sensing and the detection of charge tunneling on microsecond time scales. 
The GHz detection of dispersive resonator shifts allows rapid acquisition of charge-stability diagrams, as well as resolving charge tunneling in the device with a signal-to-noise ratio of up to 15 in one microsecond. 
Our measurements show that GHz-frequency resonators may serve as a universal tool for fast tune-up and high-fidelity readout of semiconductor qubits.
\end{abstract}

\maketitle

\section{\label{sec:introduction}Introduction}
Microwave resonators in the few-GHz range are well-known as a powerful means to increase the speed with which properties of mesoscopic quantum devices can be read out \cite{Schoelkopf_1998}.
In the field of quantum information, resonators in this so called `Super High Frequency' (SHF) band have thus enabled the fast and high-fidelity non-demolition readout of quantum bits (qubits)\cite{Blais_2004, Wallraff_2005, Vijay_2011, Walter_2017}, as well as mediating interactions between qubits \cite{Majer_2007,Sillanp__2007, Scarlino_2019, Borjans_2019}.
SHF resonators are also an attractive tool for the fast characterization of quantum devices, because the required tune-up routines are generally time-consuming.
Additionally, frequency multiplexing using many high-Q resonators has been established for hardware-efficient mass-characterization of devices \cite{Jerger_2012, Jeffrey_2014}.

Efficient characterization is particularly relevant for semiconductor quantum devices where many gate electrodes result in a large parameter space.
In recent years there have been numerous efforts to utilize SHF resonators for this purpose \cite{Puebla_Hellmann_2012, Hasler_2015, Stehlik_2015, Ranjan_2015} as well as reading out qubit degrees of freedom \cite{Delbecq_2011, Frey_2012, Petersson_2012, Mi_2016, Stockklauser_2017, Bruhat_2018, Landig_2018, Mi_2018, Samkharadze_2018, Zheng_2019, Koski_2020}.
Despite these successes, however, experiments are still often supplemented with DC or low-frequency measurements to quantitatively extract the DC conductance \cite{Harabula_2017}.
As larger-scale devices are developed \cite{Holman_2020_2, Ruffino2021_x}, it is interesting to direct focus to readout and tune-up schemes utilizing SHF resonators only, thus allowing a single framework for all measurements performed on a device.

Here, we present experiments using multiplexed resonators in the 3--7\,GHz range coupled to a multi-quantum dot (QD) system.
Using the resonator response only we are able to infer quantitatively the DC conductance of the system, and detect single-electron tunneling with high signal-to-noise ratio (SNR) on sub-microsecond timescales.
The remainder of this paper is organized as follows:
In Sec.~\ref{sec:conductance} we determine the DC (i.e., zero-frequency) conductance from SHF measurements without any DC calibration data and find agreement with conductance obtained from a DC transport control measurement.
In Sec.~\ref{sec:multiplexing}, we demonstrate fast multiplexed dispersive gate sensing (DGS) at \si{\giga\hertz}-frequencies in a double quantum dot (DQD).
This local measurement of charge transitions facilitates fast tune-up of multi-QD systems \cite{Ruffino2021_x}.
Finally, in Sec.~\ref{sec:SNR}, we attain high SNRs in the detection of charge tunneling in the DQD.
State-dependent charge-tunneling is a key mechanism for qubit readout in semiconductor qubits \cite{Derakhshan_Maman_2020}.
Our optimized resonator design \cite{Ahmed_2018}, combined with the use of a near-quantum limited amplifier \cite{Macklin_2015}, results in a maximum SNR of 15 in one microsecond integration time.

\begin{figure}[ht!]
    \centering
    \includegraphics{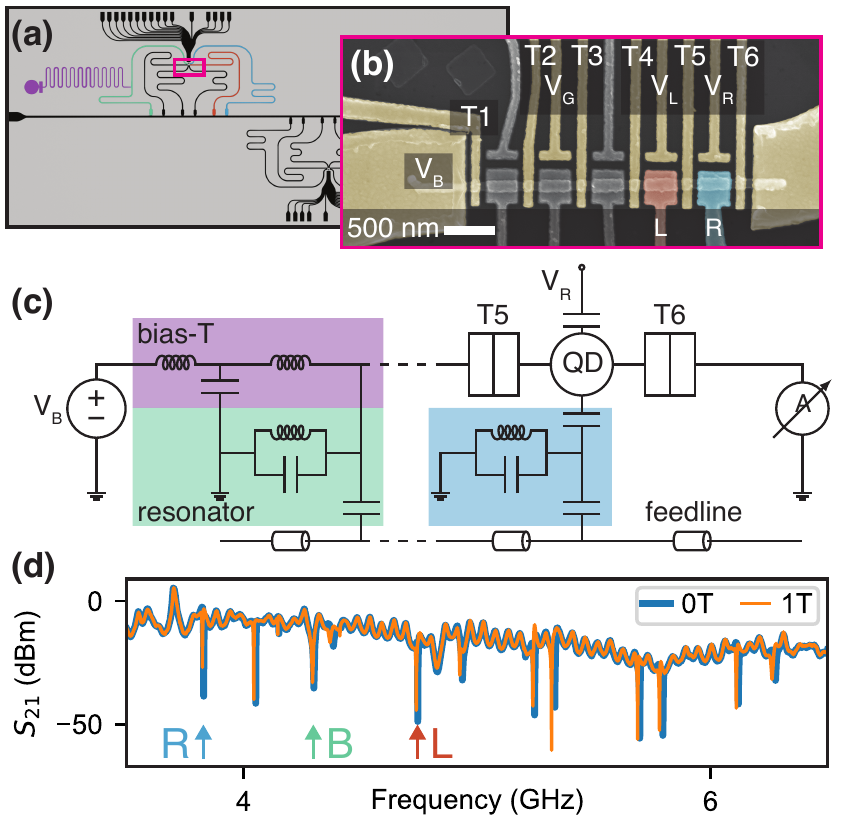}
    \caption{{\bf Experimental setup and resonator response}
        {\bf a} Schematic of the device layout {\bf b)} False-colored electron micrograph of the nanowire and the surrounding gates.
        {\bf c} RF equivalent circuit diagram of the device. The five topgates are coupled resonators as is the source of the nanowire which can be DC-biased by $V_\mathrm{B}$ with a bias-T.
        The topgates are separated by six tunnel gates such that the nanowire can be pinched off at various positions and quantum dots can be defined.
        The charge on the quantum dots can be controlled by the sidegates.
        {\bf d} Transmission through the feedline without magnetic field and at \SI{1}{\tesla} applied parallel to the plane of the resonators.
        The arrows (L)eft, (R)ight, and (B)ias mark the resonators used here.
    }
    \label{fig1}
\end{figure}

\section{Experimental setup}
The device comprises an InAs nanowire with
a GHz-bandwidth coplanar waveguide (CPW) resonator \cite{Kroll_2019} coupled to every QD to sense the electronic compressibility of each individual dot. 
An additional resonator that is galvanically connected to the source of the nanowire is used to probe the admittance of the nanowire.
Fig.~\ref{fig1}a and b show images of the resonators and the multi-QD device respectively.
An approximate lumped-element schematic of the device is shown in Fig.~\ref{fig1}c.
Each resonator is coupled to a central feedline in a hanger geometry and is individually addressable using frequency multiplexing (Fig.~\ref{fig1}d).
The obtained SNR is set by the high resonator bandwidth, optimized resonator coupling quality factors, and a traveling wave parametric amplifier (TWPA) \cite{Macklin_2015} at the base temperature stage of \SI{20}{\milli\kelvin} of our dilution refrigerator.
For further details see the supplemental material.

\section{High frequency conductance measurements}
\label{sec:conductance}
\begin{figure}[t]
    \centering
    \includegraphics[width=0.5\textwidth,trim=0 0 0 0,clip]{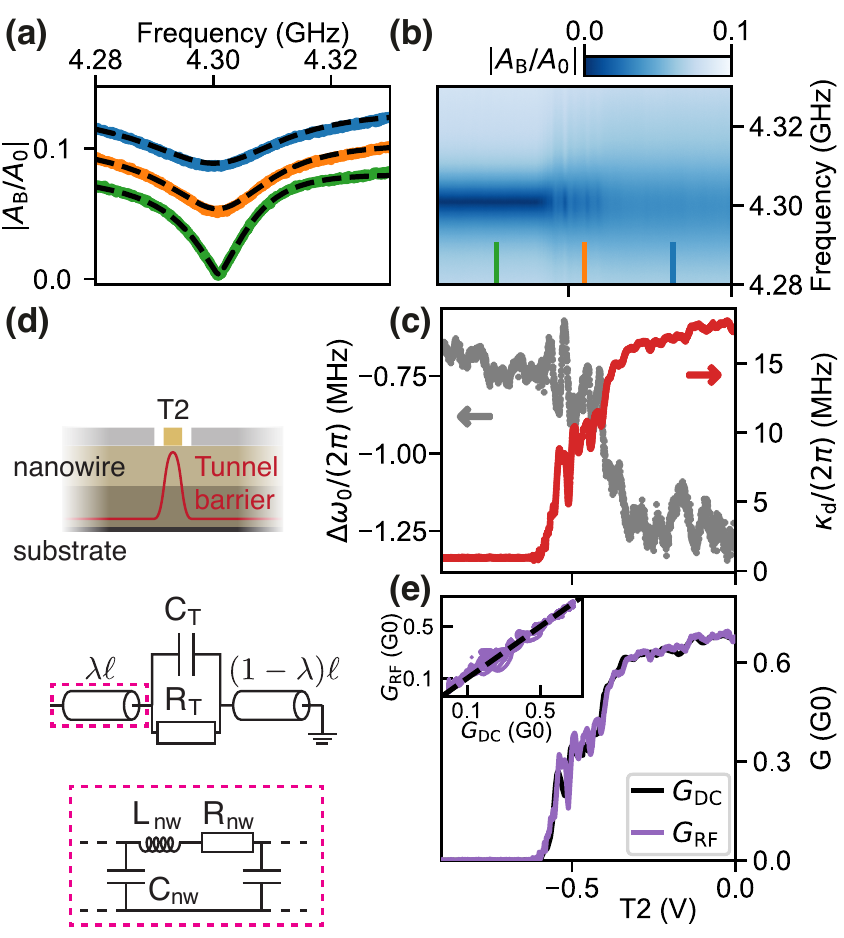}
    \caption{{\bf Pinch-off measurements.}
        {\bf a}, {\bf b} Response of the conductance resonator to the tunnel gate T2 and linecuts of the indicated gate voltages in {\bf b} offset for clarity.
        The quantity $|A_\mathrm{B}/A_0|$ denotes the ratio of measured signal to input signal.
        {\bf c} Frequency shift $\Delta \omega_0$ and internal resonator decay $\kappa_\text{d}$ extracted from individual resonator line traces of {\bf b}.
        {\bf d} Schematic of the nanowire for the experiment in {\bf b} with the corresponding lumped element model used to convert between resonator admittance and conductance $G_\text{RF}$.
        {\bf e} Conductance $G_\text{DC}$, measured with standard voltage biased current measurements, together with the conductance $G_\text{RF}$ extracted from {\bf c}.
        The inset shows the conductance $G_\text{RF}$, as a function of conductance $G_\text{DC}$, for the gate response of all tunnel gates T1 through T6.
        The dashed line indicates $G_\text{DC}=G_\text{RF}$.
        The individual traces are included in the supplemental material.
        All measurements in this figure are taken at $V_\mathrm{B}=\SI{10}{\milli\volt}$.
    }
    \label{fig2}
\end{figure}
We begin by investigating the SHF response of the resonator coupled to the lead in response to changing nanowire conductance \cite{Chorley_2012, Puebla_Hellmann_2012, Hasler_2015, Harabula_2017, Razmadze_2019}.
Changing the conductance by tuning the gate voltage T2 modulates the resonator response, shown in Fig.~\ref{fig2}a and b, through changes in its load admittance.
The DC conductance can be extracted from the load admittance either by building up a calibration map of load admittance and DC conductance, or by quantitatively modeling the resonator circuit \cite{Harabula_2017}.
We take the latter approach to maintain independence from DC calibration measurements.
To quantify the modulation of the resonator response, we fit the response to a hanger input-output model \cite{Khalil_2012, Probst_2015, Guan_2020}.
The relevant parameters for extracting load admittance are the change in the resonance frequency $\Delta\omega_0$ and the additional photon decay rate $\Delta\kappa_\mathrm{d}$ with respect to the pinched-off regime,
which is reached by decreasing the gate voltages until $\kappa_\mathrm{d}$ saturates.
Representative fits are plotted in Fig.~\ref{fig2}a and the extracted $\kappa_\mathrm{d}$ and $\Delta \omega_0$ are shown in Fig.~\ref{fig2}c.
The load admittance, $Y_{\load}$, can then be calculated by
\begin{equation}
\label{eq:yomega}
Y_{\load} = \frac{\pi}{Z_0 \omega_0}\left(\frac{1}{2}\Delta\kappa_\mathrm{d} - \imath\Delta\omega_0\right),
\end{equation}
which holds for a transmission line resonator of characteristic impedance $Z_0$ coupled to a high impedance load $1/\vert Y_{\load}\vert \gg Z_0$.
We estimate $Z_0=\SI{116}{\ohm}$ from the resonator design.
See the supplemental material for more details on the procedure outlined above.

Importantly, the load admittance at finite frequency does not directly translate to the DC conductance of the coupled device (i.e., the nanowire).
The nanowire itself has an inductive component and the gates surrounding the nanowire add additional shunting capacitive paths to ground, contributing to the load admittance especially for higher frequencies.
Our device design with high lever-arm gates necessitates compensating for these contributions explicitly, in contrast to the experiments in Refs.~\cite{Puebla_Hellmann_2012, Hasler_2015, Harabula_2017}.
To account for these effects, we model the load admittance $Y_{\load}$ as in Fig.~\ref{fig2}d, describing an effective transmission line formed by the nanowire split by a tunnel junction.
We denote the series resistance, inductance and parallel capacitance per unit length of this transmission line by $R_\mathrm{nw}, L_\mathrm{nw}$ and $C_\mathrm{nw}$ and introduce $\znw = \ell (R_\mathrm{nw}+\imath \omega L_\mathrm{nw})$ with $\ell$ the nanowire length.
DC conductance of the nanowire can be calculated from the impedance added by the nanowire itself, $\znw$, and the impedance of the tunnel junction $Z_{\mathrm{T}}$.

The relation between $Z_{\mathrm{T}}$ and $Y_{\load}$ depends on the fractional position of the tunnel junction along the nanowire, which we parameterize by $\lambda$.
Explicitly, the relation is given by
\begin{equation}
\label{eq:zt}
Z_\mathrm{T}\!=\!\frac{\frac{\znw}{\gamma \ell}}{\cosh(\!(1\!-\!\lambda)\gamma \ell)}
\frac{\frac{Y_{\load}\znw}{\gamma \ell}\!\sinh(\gamma \ell)\!-\!\cosh(\gamma \ell)}
{\sinh(\lambda \gamma \ell)\!-\!\frac{Y_{\load}\znw}{\gamma \ell}\!\cosh(\lambda \gamma \ell)},
\end{equation}
where $\gamma \equiv \sqrt{(R_\mathrm{nw}+\imath \omega L_\mathrm{nw})\imath \omega C_\mathrm{nw}}$ denotes the complex propagation constant.

The constants $\znw$ and $\gamma \ell$ are determined from two SHF calibration measurements.
For the first calibration measurement, the load impedance $Y_\mathrm{o}$ is measured when all gates are open at \SI{0}{\volt}, corresponding to the limit that $Z_\mathrm{T}=0$.
For the second calibration measurement, the load impedance $Y_\mathrm{p}$ as $\vert Z_T\vert\rightarrow\infty$ and $\lambda=1$ is measured by pinching off the rightmost gate T6.
Solving the resulting two equations for $\gamma \ell$ and $\znw$ yields
\begin{equation}
\label{eq:calibration}
\gamma \ell = \arctanh\left(\sqrt{\frac{Y_\mathrm{p}}{Y_\mathrm{o}}}\right)\text{ and }
\znw=\frac{\gamma \ell}{\sqrt{Y_\mathrm{p} Y_\mathrm{o}}}.
\end{equation}
See supplemental material for more information.
Using Eq.~\eqref{eq:zt} and Eq.~\eqref{eq:calibration}, we then extract $Z_\mathrm{T}$ from the admittance $Y$.
We model the junction as a resistor $R_\mathrm{T}$ and capacitor $C_\mathrm{T}$ in parallel such that $Z_T^{-1} \equiv 1/R_T + i\omega C_T$ \cite{van_der_Wiel_2002}, and then determine the DC-equivalent conductance as
\begin{equation}
\label{eq:grf}
G_\mathrm{RF}^{-1}=\mathrm{Re}(\znw)+ 1/\mathrm{Re}(Z_\mathrm{T}^{-1}).
\end{equation}
To validate our method to infer the conductance, we compare it to the conductance obtained from a control experiment using conventional DC-current detection.
Fig.~\ref{fig2}e shows the conductance extracted from DC measurements $G_\mathrm{DC}$ and the DC conductance extracted from the resonator response $G_\mathrm{RF}$.
Excellent agreement is observed between $G_{\mathrm{RF}}$ and $G_{\mathrm{DC}}$ for data from pinch-off traces of T1 through T6, changing $\lambda$ according to the position of the gate, shown in the inset of Fig.~\ref{fig2}e.

\begin{figure}[ht!]
    \centering
    \includegraphics[width=0.5\textwidth,trim=0 0 0 0,clip]{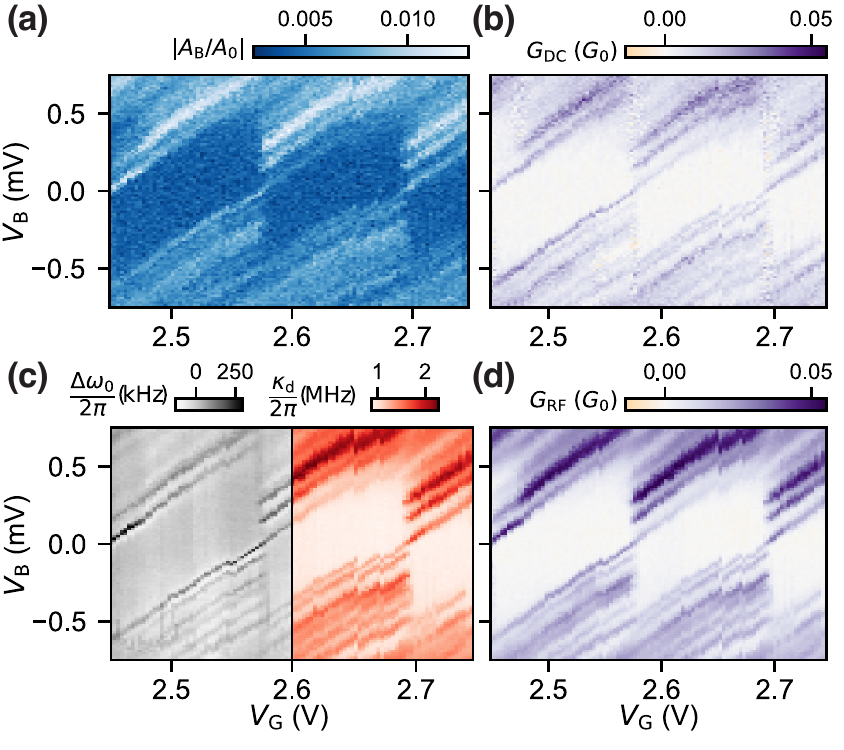}
    \caption{{\bf Coulomb blockade diamonds measured in a single quantum dot.}
        {\bf a} Single frequency response of the resonator.
        {\bf b} $G_\text{DC}$ measurements obtained with standard lock-in techniques
        {\bf c} Frequency shift $\Delta f$ and resonator decay rate $\kappa_\text{d}$ extracted from frequency traces.
        {\bf e } Conductance $G_\text{RF}$ extracted from resonator data in {\bf c}.
    }
    \label{fig3}
\end{figure}

Applications of RF conductance are not limited to measuring the impedance of tunnel gates \cite{Gabelli2006_x, Jung_2012, Ranjan_2015, Ares_2016}.
As an example, we probe a quantum dot by tuning T2 and T3 into the tunneling regime.
We show the amplitude response of the lead resonator on resonance in Fig.~\ref{fig3}a as a function of bias voltage $V_\mathrm{B}$ and gate voltage $V_\mathrm{G}$.
Even though the amplitude response is not translated into DC conductance here, it shows all the qualitative features present in the control data measured by DC lock-in conductance (Fig.~\ref{fig3}b), including the excited states of the quantum dot.
The amplitude response of Fig.~\ref{fig3}a is part of a full frequency trace, measured to also allow for a quantitative comparison between the DC results and the resonator response.
From these traces, the frequency shift $\Delta\omega_0$ and photon decay $\Delta\kappa_\mathrm{d}$ are extracted and shown in Fig.~\ref{fig3}c.
We use the model defined by Eqs.~(\ref{eq:yomega}~-~\ref{eq:grf})
to obtain $G_\mathrm{RF}$, shown in Fig.~\ref{fig3}d.
This is the same model used for the tunnel junction scans of Fig.~\ref{fig2}.
Note that we neglect here the finite width occupied by the quantum dot and its internal structure; nevertheless we observe reasonable agreement between $G_\mathrm{RF}$ and $G_\mathrm{DC}$.

\section{Rapid multiplexed reflectometry}
\label{sec:multiplexing}
\begin{figure}[t!]
    \centering
    \includegraphics[width=0.5\textwidth,trim=0 0 0 0,clip]{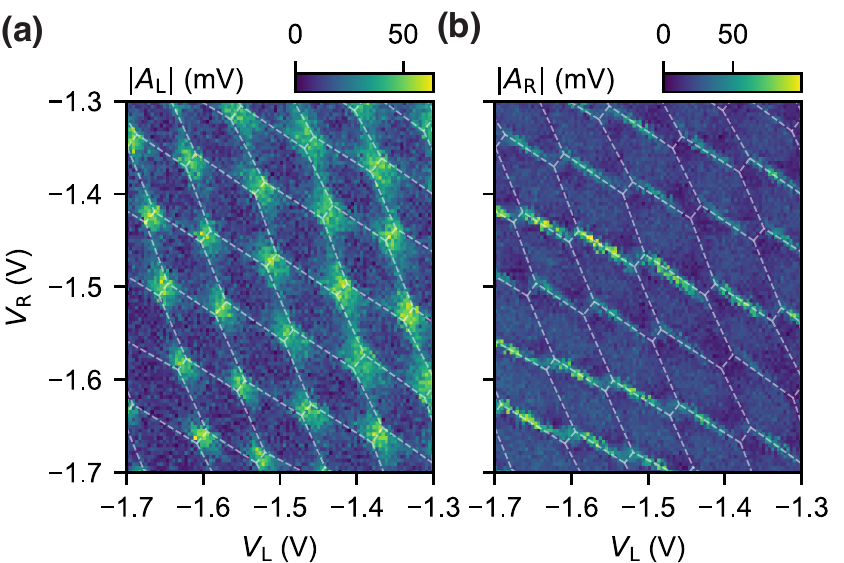}
    \caption{{\bf Charge stability diagram measured using multiplexed gate-based readout in the double dot regime.}
        {\bf a}, {\bf b} Amplitude response of the resonators coupled to the two rightmost quantum dots.
        Readout power in the feedline is \SI{-105}{\dBm} per multiplexed resonator with an integration time of \SI{3}{\micro\second}.
        The dimensions of this dataset are $101\times101$ points yielding a total integration time of \SI{30}{\milli\second} excluding overhead from gate settling time, set by low pass filters on the gate wiring.
        The dashed lines are guides to the eye delineating the different charge configurations of the double dot and are identical on the left and right plots.}
    \label{fig4}
\end{figure}

\begin{figure*}[t!]
	\centering
	\includegraphics[width=1\textwidth,trim=0 0 0 0,clip]{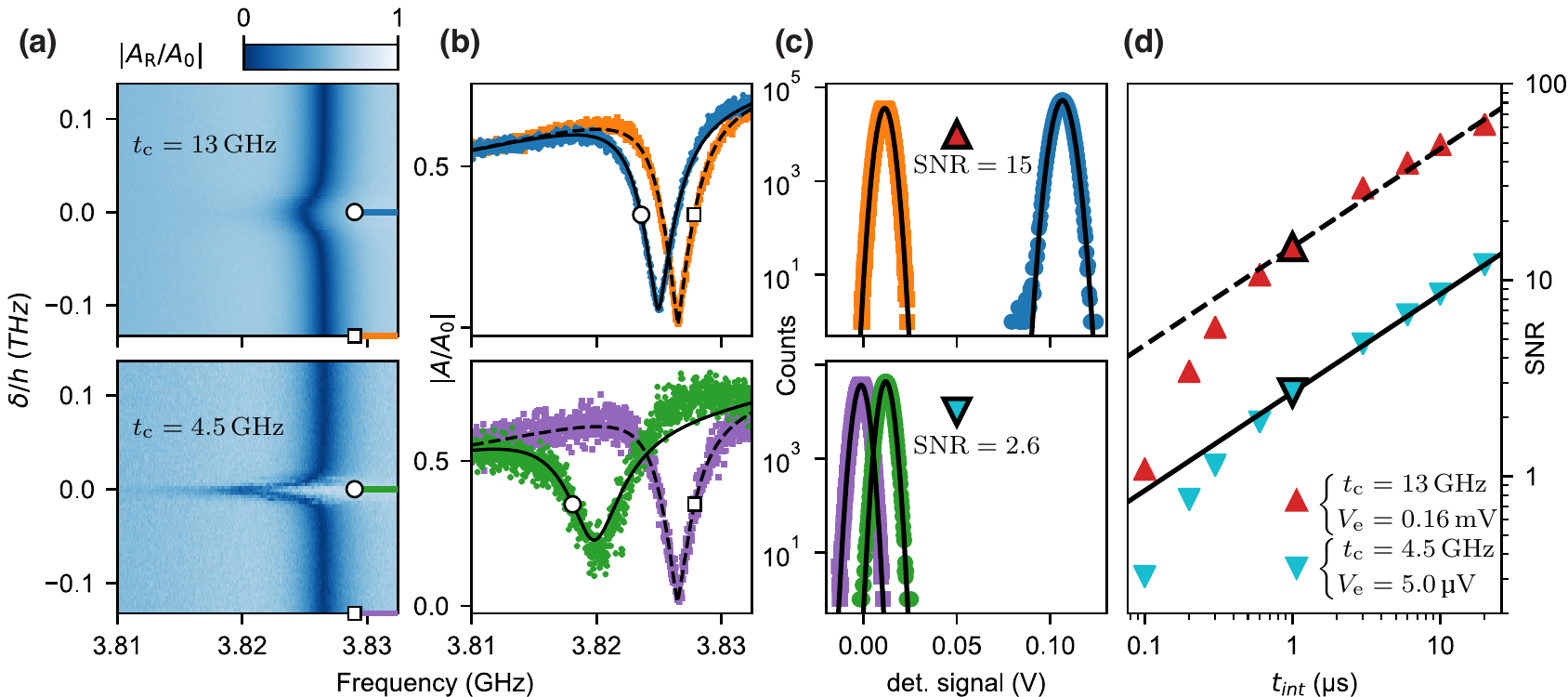}
	\caption{{\bf Readout SNR.}
        {\bf a}, Amplitude response as a function of detuning $\delta$ of the resonator coupled to the right dot for the two different tunnel coupling regimes.
        Linecuts are shown in respectively in {\bf b} for Coulomb blockade (square marker) and on resonance (circle marker) together with fits to the theoretical model.
        {\bf c}, Histograms of the resonator responses in Coulomb blockade and charge degeneracy, with pulse length of \SI{1}{\micro\second}.
        Responses were acquired with a probe frequency tuned to resonance for the Coulomb blockade case, at approx.~3.826\,GHz.
		{\bf d} Attained SNR on the right dot's resonator, defined as $\Delta / (2 \sigma)$, as a function of measurement pulse length. 
		Optimized with excitation voltage as a free parameter (red, cross markers) and optimized at fixed excitation voltage of $\SI{5}{\micro\volt}$ (aqua, star markers).
	}
	\label{fig5}
\end{figure*}

We now move on to the capacitively coupled gate resonators and investigate DGS in the double quantum dot (DQD) regime \cite{Petersson_2010, Frey_2012, Colless_2013, Betz_2015, Esmail_2017,  Pakkiam_2018, Urdampilleta_2019, West_2019, Zheng_2019, de_Jong_2019, Sabonis_2019, Crippa_2019}.
To tune the system into a DQD, the voltages of gates T4, T5, and T6 are each decreased into the tunneling regime.
Accordingly, two quantum dots are formed under the rightmost two topgates in the nanowire\cite{van_der_Wiel_2002}.

A resonator is coupled to both dots to sense the electronic compressibility of the individual dots  \cite{Esterli_2019,Park_2020}.
In Fig.~\ref{fig4} we show a charge stability diagram (CSD) using $V_\mathrm{L}$ and $V_\mathrm{R}$ to change the electron occupation of the DQD.
We perform pulsed readout with an integration time of \SI{3}{\micro\second} per point, constituting a total data acquisition time of \SI{30}{\milli\second} for the entire CSD \cite{Stehlik_2015, Schupp_2020}.
The data acquisition is frequency-multiplexed for both resonators such that the data in panels a and b of Fig.~\ref{fig4} are measured simultaneously \cite{Hornibrook_2014,Ruffino2021_x}.
This not only reduces the measurement time, but multiplexing also guarantees the measurements in Fig.~\ref{fig4}a and b correspond to the exact same physical regime, regardless of charge jumps and gate hysteresis. 
To emphasize the correspondence between Fig.~\ref{fig4}a and b, the same guides to the eye outlining stable charge configurations are drawn on both figures.

Resonators are only sensitive to charge transitions involving the quantum dots they are coupled to.
Therefore, both resonators show the interdot transitions, but only the left/right resonator is sensitive to transitions to the left/right lead respectively.
Hence multiplexing also enables spatial correlation of electron tunneling by comparing the DGS signal from each gate’s resonator, effectively `tracking' the electron through the device.

\section{Signal to noise}
\label{sec:SNR}
Finally, we investigate the attainable SNR for resolving charge tunneling with DGS
by changing detuning from charge degeneracy in the DQD. 
This procedure serves as a proxy for different qubit states in schemes where readout is based on state-dependent tunneling \cite{Plugge_2017, Karzig_2017, de_Jong_2019, Razmadze_2019, Zheng_2019, Smith_2020}.
Because actual qubit systems will have limitations on the readout power \cite{Derakhshan_Maman_2020} we investigate the SNR at both a fixed `low' excitation voltage in the resonator, $V_\mathrm{e}=\SI{5}{\micro\volt}$, as well as at an optimized excitation voltage, $V_\mathrm{e} = \SI{0.16}{\milli\volt}$.
These excitation voltages are calculated from the generator power and line attenuation in addition to the resonator frequency and coupling capacitance to the feedline.

We fix the total charge in the system by pinching off gates on either side of the DQD.
The only remaining transitions are interdot transitions occurring through a tunnel coupling denoted by $t_\mathrm{C}$.
The resonator response as a function of the energy detuning $\delta$ from the interdot transition, is shown in Fig.~\ref{fig5}a.
We determine $t_\mathrm{C}$ by fitting the resonator response to an input-output model \cite{Petersson_2012}, discussed in the supplemental material.
Linecuts of the fit results and measurement data are shown in Fig.~\ref{fig5}b.

We define SNR as the change in signal between charge degeneracy and Coulomb blockade divided by the noise.
To measure it, we perform a series of pulsed measurements of I and Q with a pulse time of $t_\mathrm{m}$ at both Coulomb blockade and charge degeneracy, and show the obtained histograms for an integration time of $t_\mathrm{m}=\SI{1}{\micro\second}$ in Fig.~\ref{fig5}c.
These histograms are fit with a Gaussian to extract the separation between the Gaussian peaks $\Delta$ in the IQ plane as well as their average standard deviation $\sigma$ representing the width.
The SNR is given by $\Delta / (2 \sigma)$.

In Fig.~\ref{fig5}d we plot the dependence of SNR on $t_\mathrm{m}$, which approaches a square root dependence for longer times.
We attribute the discrepancy between attained SNR and a square root dependence for pulse times shorter than \SI{1}{\micro\second}
to the finite bandwidth of the resonators. 
For these pulse lengths, the resonator cannot reach a steady state photon population, limiting the signal available for readout.

Next, we compare the observed SNR with expected theoretical limits.
The change in signal at the feedline level $\Delta_\mathrm{f} = \Delta/G_\mathrm{sys}$ --- with $G_\mathrm{sys}$ the gain of the amplification chain in the system --- can never exceed the total voltage swing in the feedline $V_\mathrm{f}$.
The fit to the data in Fig.~\ref{fig5}a, used to extract $t_\mathrm{C}$, also provides a direct measurement of the ratio $\Delta_f/V_\mathrm{f} = 0.89$, close to the absolute maximum.
In other words, the resonator is coupled near-optimally for this tunnel coupling, such that its external coupling rate is nearly equal to the dispersive shift.

The achievable SNR together is then set by $\Delta_\mathrm{f}$ together with noise temperature, $T_\mathrm{N}$, and readout time, $t_\mathrm{m}$, as:
\begin{equation}
\mathrm{SNR} = \frac{\Delta_\mathrm{f}\sqrt{t_\mathrm{m}}}{2\sqrt{Z k_\mathrm{B} T_\mathrm{N}}},
\end{equation}
where $Z=\SI{50}{\ohm}$ is the impedance of the feedline, see supplemental material.
The $\mathrm{SNR}\simeq2.6$ found in Fig.~\ref{fig5}c together with the readout time, $t_\mathrm{m}=\SI{1}{\micro\second}$ and the deduced approximate voltage swing in the feedline, $V_\mathrm{f}=\SI{0.15}{\micro\volt}$ 
corresponds to a noise temperature estimate of $T_\mathrm{N}=\SI{1}{\kelvin}$.
To improve the SNR, one can either increase the readout time or readout power, as shown in Fig.~\ref{fig5}d.
In practice, limits to these two parameters will be determined by the specific qubit implementation.
Specifically, by optimizing the excitation voltage and tunnel coupling together, an $\mathrm{SNR}$ of $15$ is achieved at $V_\mathrm{e}=\SI{0.16}{\milli\volt}$.

\section{\label{sec:conclusions}Conclusions}
We have shown the characterization of an InAs nanowire multi-QD system using \si{\giga\hertz}-frequency sensing.
Probing the finite frequency admittance of the nanowire has allowed us to infer the low-frequency conductance with good accuracy, even without calibration from DC measurements.
Further, we have shown high-SNR dispersive sensing on timescales near the bandwidth limit set by the Q factor of the resonators.
Besides their use for qubit devices, we envision that fast multiplexed readout of quantum devices may be used for more complex sensing schemes.
In particular, rapid simultaneous conduction of multiple local measurements could facilitate unique quantum transport experiments because they provide spatial information about tunneling processes.
For example, probing two quantum dots at either end of a central charge island, tunneling events into the outer dots may be correlated \cite{Tan_2015, M_nard_2020}.
We conclude that multiplexed SHF resonators may serve as a complete toolset for characterization and readout of semiconductor quantum devices, and present intriguing opportunities for developing novel and high-speed quantum transport measurement schemes.

\section{\label{sec:acknowledgements}Acknowledgements}
We thank D. Bouman and J.D. Mensingh for nanowire deposition and A. Bargerbos for valuable comments on the manuscript.
We further thank N.P. Alberts, O.W.B. Benningshof, R.N. Schouten, M.J.Tiggelman, and R.F.L. Vermeulen for valuable technical assistance.
This work has been supported by the Netherlands Organization for Scientific Research (NWO) and Microsoft.

\end{document}


\title{Supplementary information for ``Rapid microwave-only characterization and readout of quantum dots using multiplexed gigahertz-frequency resonators"}

\author{Damaz de Jong}
\affiliation{QuTech and Kavli Institute of Nanoscience, Delft University of Technology, 2600 GA Delft, The Netherlands}

\author{Christian Prosko}
\affiliation{QuTech and Kavli Institute of Nanoscience, Delft University of Technology, 2600 GA Delft, The Netherlands}

\author{Daan M. A. Waardenburg}
\affiliation{QuTech and Kavli Institute of Nanoscience, Delft University of Technology, 2600 GA Delft, The Netherlands}

\author{Lin Han}
\affiliation{QuTech and Kavli Institute of Nanoscience, Delft University of Technology, 2600 GA Delft, The Netherlands}

\author{Filip K. Malinowski}
\affiliation{QuTech and Kavli Institute of Nanoscience, Delft University of Technology, 2600 GA Delft, The Netherlands}

\author{Peter Krogstrup}
\affiliation{Center for Quantum Devices, Niels Bohr Institute, University of Copenhagen \& Microsoft Quantum Materials Lab Copenhagen, Denmark}

\author{Leo P. Kouwenhoven}
\affiliation{QuTech and Kavli Institute of Nanoscience, Delft University of Technology, 2600 GA Delft, The Netherlands}
\affiliation{Microsoft Quantum Lab Delft, Delft University of Technology, 2600 GA Delft, The Netherlands}
\author{Jonne V. Koski}
\affiliation{Microsoft Quantum Lab Delft, Delft University of Technology, 2600 GA Delft, The Netherlands}
\author{Wolfgang Pfaff}
\affiliation{Microsoft Quantum Lab Delft, Delft University of Technology, 2600 GA Delft, The Netherlands}
\affiliation{Department of Physics and Frederick Seitz Materials Research Laboratory, University of Illinois at Urbana-Champaign, Urbana, IL 61801, USA}

\date{\today}
\maketitle

\renewcommand\theequation{S\arabic{equation}}
\renewcommand\thefigure{S\arabic{figure}}
\renewcommand\thetable{S\arabic{table}}

\section{Sample fabrication and experimental setup}

\begin{figure}[ht!]
    \includegraphics{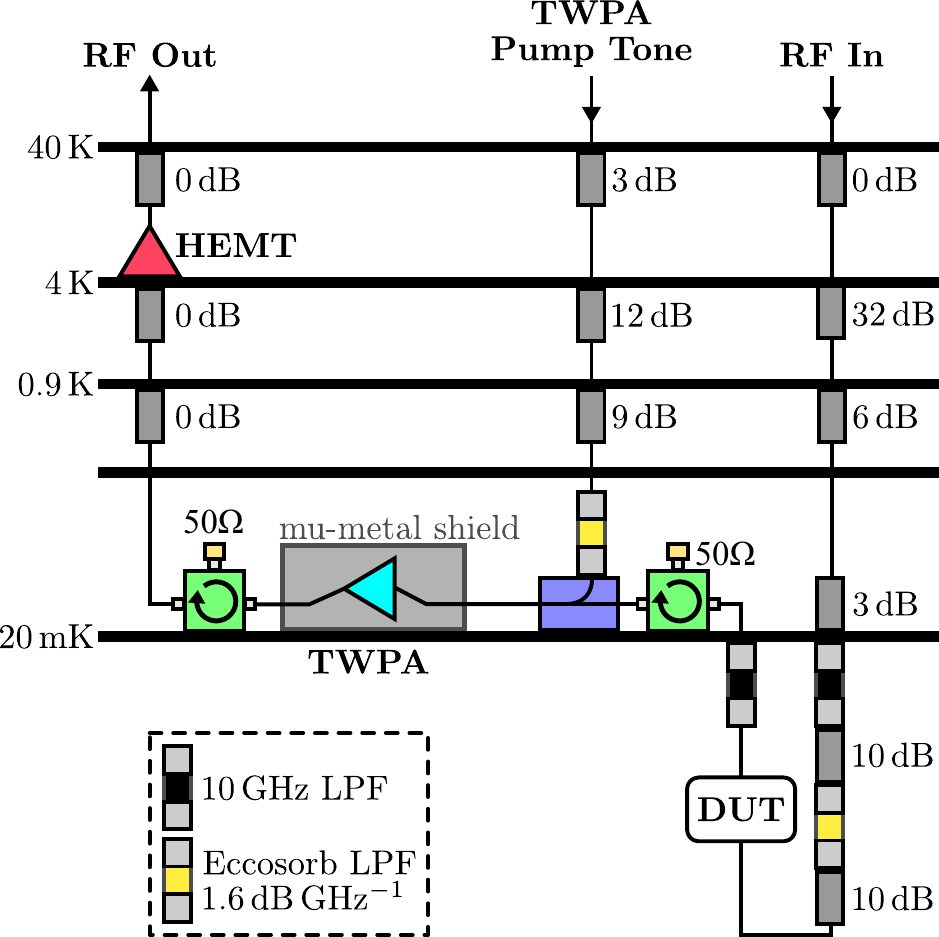}
        \caption{
            {\bf RF circuit of the dilution refrigerator.}
            An RF signal generated either by a vector network analyzer or Quantum Machines OPX pulse generator is attenuated at each stage of the refrigerator on its way to the device under test (DUT) at base temperature.
            The signal is amplified on the way out of the fridge first by a traveling wave parametric amplifier (TWPA) at base temperature, driven by a \si{\giga\hertz}-frequency pump tone, and then by a 4 to \SI{8}{\giga\hertz} bandwidth high electron mobility transistor (HEMT) amplifier at the \SI{4}{\kelvin} plate.
            In front of and behind the TWPA are circulators (green) and a directional coupler (violet) designed to reduce back-action of the TWPA on the DUT and attenuate any noise traveling down the output and pump tone lines.
            Low pass filters are also present to reduce noise above the measurement bandwidth.
        }
        \label{fig:fridge_circuit_suppl}
\end{figure}

The experimental setup is schematically shown in Fig.~\ref{fig:fridge_circuit_suppl}.
On-chip superconducting coplanar waveguide (CPW) resonators are fabricated from a \SI{20}{\nano\meter} NbTiN layer using reactive ion etching, similar to Ref.~\cite{Kroll_2019} ensuring magnetic field compatibility.
The resonator design is half-wavelength with hanger-style geometry, with each resonator coupled to a central feedline.
The resonators used for DGS have an external coupling factor around $\kappa_\mathrm{ext}\simeq\SI{10}{\mega\hertz}$ to maximize readout signal.

An InAs nanowire with an epitaxial Al shell is deposited using a micromanipulator.
The superconducting shell is removed everywhere except over the middle dot using a PMMA etch mask and a \SI{70}{\second} wet etch using MF-321 developer.
However, no superconducting effects are observed in this experiment.
We attribute the absence of superconductivity to over-etching of the Al shell on the wire.
Contacts are fabricated by in-situ argon milling followed by evaporating \SI{10}{\nano\meter} Ti then \SI{150}{\nano\meter} Au.
A \SI{10}{\nano\meter} AlOx dielectric is selectively deposited on the nanowire, away from the resonators to avoid additional dielectric loss.
Lastly, top gates are evaporated in the same way as the contacts.
To minimize the noise temperature in our measurements we use a TWPA \cite{Macklin_2015} on the base temperature stage of a dilution refrigerator operating at \SI{20}{\milli\kelvin}.
Additionally, a high electron mobility transistor amplifier at \SI{4}{\kelvin} is used to further amplify the signal.

\section{Fitting procedure for resonator response}

In this section we describe the complete fitting procedure for the measured resonator response.
The model used for fitting the transmission $S_{21}$ through a feedline with a hanger-style resonator is given by\cite{Khalil_2012, Guan_2020, Probst_2015}:
\begin{equation}
\label{eq:s21}
S_{21} = e^{\imath (\theta_0+\theta_1 \omega)}  s_0 \left(1+s_1\frac{\omega-\omega_0}{\omega_0}\right)
\left( 1 - \frac{1}{2}\frac{\imath \kappa_\mathrm{ext}}{\imath (\kappa_\mathrm{ext}+\kappa_\mathrm{d})/2-\omega+\omega_0} \right).
\end{equation}
The probe frequency is denoted by $\omega$ and the resonator frequency by $\omega_0$.

To account for the line delay, $\theta_0$ and $\theta_1$ account for a linear offset in phase.
Similarly, we account for a linear amplitude offset by $s_0$ and $s_1$.
The photon decay is represented by $\kappa_\mathrm{d}$ and the external coupling factor $\kappa_\mathrm{ext}$ is complex to account for impedance mismatches.

The calibration frequency trace used for the data in Fig.~2 of the main paper is shown in Fig.~\ref{fig:suppl_pinchoff}a together with a fit to Eq.~\eqref{eq:s21}.
We find $\omega_0/2\pi = \SI{4.3}{\giga\hertz}$, $\kappa_\mathrm{ext}/2\pi=(18.6-3.4\imath)\,\mathrm{MHz}$ and $\kappa_\mathrm{d}/2\pi = \SI{0.9}{\mega\hertz}$.
The calibration measurement defines the zero-point of $\Delta\omega_0$ and $\Delta\kappa_\mathrm{i}$.
We also use the calibration to hold all parameters except for $\omega_0$ and $\kappa_{\mathrm{i}}$ fixed when fitting the frequency traces for obtaining the pinch-off data, such as in Fig.~2b.

For fitting the dispersive shift as a function of detuning $\delta$, in Fig.~5 of the main paper we add the contribution of the DQD to Eq.~\eqref{eq:s21}.
This contribution is accounted for by substituting $\kappa_{\mathrm{d}} \rightarrow \kappa_{\mathrm{d}} - 2\imath g\chi$ in eq.~\eqref{eq:s21} caused by coupling to the susceptibility of the DQD \cite{Petersson_2012} with
\begin{equation}
g\chi = \frac{4 g_0^2t_\text{C}^2/\Omega^2}{\omega_0-\Omega+\imath \gamma/2},
\end{equation}
where $g$ is the effective coupling strength and $\chi$ the susceptibility of the DQD.
Furthermore, $g_0$ is the Jaynes-Cummings coupling, $\Omega=\sqrt{4t_\text{C}^2+\delta^2}$ is the DQD energy splitting, $t_\text{C}$ the tunnel coupling between the dots and $\gamma$ the decoherence rate.

\section{Correspondence between $\loadadm$ and resonator response}

Here we relate the nanowire load admittance to the quality factor and resonance frequency of a resonator connected to its lead.
Throughout the following derivations, we assume that within the small window of frequencies used to fit our resonator's resonance frequency and quality factor, the sample load admittance $\loadadm$ is constant.
The resonator is modeled as a transmission line capacitively coupled to a central feedline and terminated by a load impedance $\loadimp$ determined by the nanowire.
Assuming the feedline coupler to behave as a lumped element capacitance $\couplingcap$, the input impedance $Z_\mathrm{in}$ of the terminated resonator and coupler is \cite{Pozar2011}:
\begin{equation}\label{eq:input_imp}
    \inputimp = \frac{1}{\imath\omega \couplingcap} + Z_0\frac{1 + Z_0\loadadm\tanh(\gammares \length)}{Z_0\loadadm + \tanh(\gammares \length)},
\end{equation}
where $Z_0$ is the line's characteristic impedance, $\gammares \equiv \diss + \imath\betares$ is the complex propagation constant, and $\length$ is the length of the resonator.
Note that $\betares = \omega/v_p$ where $v_p$ is the phase velocity in the transmission line, while $\diss$ quantifies internal losses.
To simplify the above expression, we first note the trigonometric relation
\begin{equation}\label{eq:trig_identity}
    \tanh(\diss\length+\imath\betares\length)
    = \frac{\sinh(\diss\length)\cos{(\betares\length)} + \imath\cosh(\diss\length)\sin(\betares\length)}{\cosh(\diss\length)\cos(\betares\length) + \imath\sinh(\diss\length)\sin(\betares\length)}
    \sim \frac{\diss\length\cos(\betares\length) + \imath\sin(\betares\length)}{\cos(\betares\length) + \imath\diss\length\sin(\betares\length)}
\end{equation}
where we have assumed small internal losses in the resonator, $\diss\length\ll 1$.
Since the load admittance is assumed to be small, our resonator nearly has an open at one end. Consequently, the effect of $\loadadm$ should be that of a perturbed $\lambda/2$ resonator. 
In this case, for small detuning $\delta\omega$ from the resonance frequency $\omega_0$, $\betares\length \sim \pi + \pi\delta\omega/\omega_0$ \cite{Pozar2011}. 
Since $Y$ is a small perturbation of the load admittance away from zero, it will shift the resonance frequency only slightly, in which case it is still true that $\betares\length \sim \pi + x $ where $x$ is a small number.
Applying this approximation to eq.~\ref{eq:trig_identity}, we find $\tanh(\gammares\length) \sim \diss\length + \imath\tan(\betares\length)$.
Finally, we assume $\omega$ is near resonance such that we may apply the limit $\tan(\betares\length)\ll 1$ by the above argument, which in combination with our assumption of small load admittance $\loadadm \ll Z_0^{-1}$ and losses $\diss\length\ll 1$ simplifies eq.~\ref{eq:input_imp} to:
\begin{equation}\label{eq:input_imp_approx}
    \inputimp \sim \frac{1}{\imath\omega \couplingcap} + \frac{Z_0}{Z_0\loadadm + \diss\length + \imath\tan(\betares\length)}
    = \frac{1}{\imath\omega \couplingcap} + \frac{Z_0}{\disseff\length + \imath\left[\tan(\betares\length) + Z_0\mathrm{Im}(Y)\right]}
\end{equation}
to first order in these small parameters. Above, we defined the effective dissipation constant $\disseff\length \equiv \diss\length + Z_0\mathrm{Re}(Y)$. 

Next, we determine the relation between $\mathrm{Im}(Y)$ and the resonance frequency $\omega_0$.
At resonance, the imaginary part of $\inputimp$ disappears, so we solve this condition for $\omega_0$:
\begin{equation}
    0 = \mathrm{Im}(\inputimp)
    = -\frac{1}{\omega_0\couplingcap} - Z_0\frac{\tan(\omega_0\length/v_\mathrm{p}) + Z_0\mathrm{Im}(Y)}{(\disseff\length)^2 + (\tan(\omega_0\length/v_\mathrm{p}) + Z_0\mathrm{Im}(Y))^2}.
\end{equation}
With the foresight that internal quality factors of our resonators will be related to $\disseff\length$ through $\disseff\length = \pi/(2Q_i)$, from resonator fits we may estimate that $\disseff\length < 0.002$ even when the nanowire is completely open. 
In our resonator chip, coupling capacitances are on the order of $\SI{40}{\femto\farad}$, so that at few-\si{\giga\hertz} frequencies and when $Z_0 = \SI{116}{\ohm}$, $\omega_o\couplingcap Z_0 \approx 0.03\text{ to }0.1$ is a small parameter, but still much larger than $\disseff\length$.
Rearranging and neglecting terms above first order in $\disseff\length/(Z_0\omega\couplingcap)$, we obtain the implicit solution:
\begin{equation}
    \frac{\omega_0\length}{\phasevel } = n\pi-\arctan\left(Z_0\mathrm{Im}(Y) + \omega_0\couplingcap Z_0\right),\,\,\,n\in\mathbb{Z}
\end{equation}
The smallest substantial resonance frequency occurs for $n=1$, constituting the \si{\giga\hertz}-range resonances of interest. 
Taylor expanding in the small parameters $Z_0\mathrm{Im}(Y)$ and $\omega_0C_cZ_0$, we see then that the coupling capacitance serves only to impose a constant perturbation to the bare resonance frequency, defined as $\barefreq \equiv \omega_0\vert_{\mathrm{Im}(Y)=0}=\pi(\length/\phasevel + \couplingcap Z_0)^{-1}$:
\begin{equation}
    \omega_0 = \omega_0^*\left(1 - \frac{Z_0\mathrm{Im}(Y)}{\pi}\right).
\end{equation}
At frequencies near resonance such that $\omega = \omega_0 + \delta\omega$ with $\delta\omega \ll \omega_0$, eq.~\ref{eq:input_imp_approx} is asymptotic to:
\begin{equation}
    \inputimp \sim \frac{1}{\imath\omega\couplingcap} + \frac{Z_0}{\disseff\length + \imath\pi\delta\omega/\omega_0},\hspace{1cm}\delta\omega \ll \omega_0,\,\disseff\length\ll Z_0\omega_0\couplingcap\ll 1,
\end{equation}
which is the input impedance of a capacitively coupled parallel $LRC$ resonator circuit near resonance of internal quality factor $Q_i = \pi/(2\disseff\length)$ \cite{Pozar2011}. The internal quality factor is related to the photon decay rate by $\kappa_\mathrm{d} \equiv \omega_0/Q_i = 2\omega_0\disseff\length/\pi$.
From the definitions of $\disseff\length$ and $\omega_0$, we can thus relate the device admittance to resonator parameters through:
\begin{equation}
    \loadadm 
    = \frac{1}{Z_0}\Big(\disseff\length - \diss\length\Big) - \imath\frac{\pi}{Z_0\omega_0}\Big(\omega_0-\omega_0^*\Big)
    \equiv \frac{\pi}{Z_0\omega_0}\left(\frac{1}{2}\Delta\kappa_\mathrm{d} - \imath\Delta\omega\right),
\end{equation}
valid to first order in $\Delta\omega/\omega_0$, where $\Delta\omega \equiv \omega_0-\omega_0^*$ and $\Delta\kappa_\mathrm{d} \equiv \kappa_\mathrm{d} - 2\omega_0\diss\length/\pi$. 
In other words, load conductance is proportional to shifts in the resonator's internal decay factor, while its susceptance is proportional to shifts in the resonance frequency. 

\section{Derivation of $G_{RF}$}

Since the nanowire device is covered at most points by a capacitively coupled gate layer of uniform thickness (excluding the small gaps between gates), we model the nanowire as a highly resistive transmission line, and aim to solve for its admittance $\loadadm$. 
As per the lumped element model of Fig.~2d, we parameterize this with a resistance, inductance, and capacitance per unit length of $R_\mathrm{nw}$, $L_\mathrm{nw}$, and $C_\mathrm{nw}$, respectively. 
At a fraction $\lenfrac$ along the wire's length $\ell$, we include a lumped element impedance $\tunnelimp$, modeling a cutter gate or quantum dot.

As a transmission line, on either side of $\tunnelimp$ the nanowire obeys the telegrapher equations \cite{Pozar2011}:
\begin{equation}
    \frac{\diff V(x)}{\diff x} = -\wireimp I(x)/\ell
    \,\text{ and }\,
    \frac{\diff I(x)}{\diff x} = -\imath\omega C_\mathrm{nw}V(x),
\end{equation}
at every point $x$ along the wire's length, with $x=0$ denoting the source lead. Above, we have assumed phasor solutions of the voltage $v$ with respect to ground and current $i$ through the wire so that $v(x, t) = V(x)e^{i\omega t}$ and $i(x, t) = I(x)e^{\imath\omega t}$. On either side of the impedance $\tunnelimp$, these coupled differential equations have the solution:
\begin{equation}\label{eq:voltage_current}
    V(x) = \left\{\begin{array}{ll}
        V^+_se^{-\gamma x} + V^-_se^{\gamma x} & x < \lenfrac l \\
        V^+_de^{-\gamma x} + V^-_de^{\gamma x} & x > \lenfrac l
    \end{array}\right.,\,\,
    I(x) = \left\{\begin{array}{ll}
        \frac{\gamma \ell}{\wireimp }\left(V^+_se^{-\gamma x} - V^-_se^{\gamma x}\right) & x < \lenfrac\ell \\
        \frac{\gamma \ell}{\wireimp }\left(V^+_de^{-\gamma x} - V^-_de^{\gamma x}\right) & x > \lenfrac \ell
    \end{array}\right.
\end{equation}
The nanowire's input admittance is $\loadadm = I(0)/V(0)$ and is fully determined by the boundary condition of a grounded wire $V(\ell) = 0$, current continuity just before and after $\tunnelimp$, and Ohm's law across $\tunnelimp$.
Combined, these three conditions allow us to solve for all constants $V_s^-$, $V_d^-$, and $V_d^+$ in terms of $V_s^+$. In particular:
\begin{equation}
    V_s^- = V_s^+\left[\frac{\frac{\tunnelimp\gamma \ell}{\wireimp }\left(e^{-2\gamma\lenfrac\ell} + e^{-2\gamma \ell}\right) -2e^{-2\gamma \ell}}{\frac{\tunnelimp\gamma \ell}{\wireimp }\left(1 + e^{-2\gamma(1-\lenfrac)\ell}\right) + 2}\right]
\end{equation}
After substituting eq.~\ref{eq:voltage_current} into the definition of $Y$, we arrive at the expression
\begin{equation}\label{eq:loadAdmittance}
    \loadadm
    = \frac{\gamma \ell}{\wireimp }\left(\frac{V_s^+ - V_s^-}{V_s^++V_s^-}\right)
    = \frac{\gamma \ell}{\wireimp }\left[\frac{\cosh(\gamma \ell) + \frac{\tunnelimp\gamma \ell}{\wireimp }\sinh(\gamma\lenfrac \ell)\cosh(\gamma(1-\lenfrac)\ell)}{\sinh(\gamma \ell) + \frac{\tunnelimp\gamma \ell}{\wireimp }\cosh(\gamma\lenfrac \ell)\cosh(\gamma(1-\lenfrac)\ell)}\right].
\end{equation}
Finally, this expression may be rearranged to yield eq.~2 of the main article.
Together, eq.~2 and eq.~4 of the main article yield an explicit formula for $G_\mathrm{RF}$.
Substituting the $\tunnelimp$ result into eq.~4 we obtain:
\begin{equation}
    G_{RF} = \frac{\vert z_\lenfrac\vert\cos(\arg{[z_\lenfrac] - \arg{[z_1]})}}{\mathrm{Re}(\wireimp )\vert z_\lenfrac\vert\cos(\arg{[z_\lenfrac]} - \arg{[z_1]}) + \vert z_1\vert}.
\end{equation}
Above, the parameters 
\begin{equation}
    z_1 \equiv \sinh(\gamma \ell)\loadadm - (\gamma \ell/\wireimp )\cosh(\gamma \ell),\text{ and } 
    z_\lenfrac \equiv (\gamma \ell/\wireimp )\cosh(\gamma(1-\lenfrac)\ell)\left[(\gamma \ell/\wireimp )\sinh(\gamma\lenfrac \ell) - \cosh(\gamma\lenfrac \ell)\loadadm\right]
\end{equation}
represent singularities of $\tunnelimp^{-1}$ and $\tunnelimp$ respectively.

\section{Determination of $\gamma \ell$ and $Z_\mathrm{nw}$ from experimental data}

\begin{figure}[ht!]
    \includegraphics[width=1\textwidth]{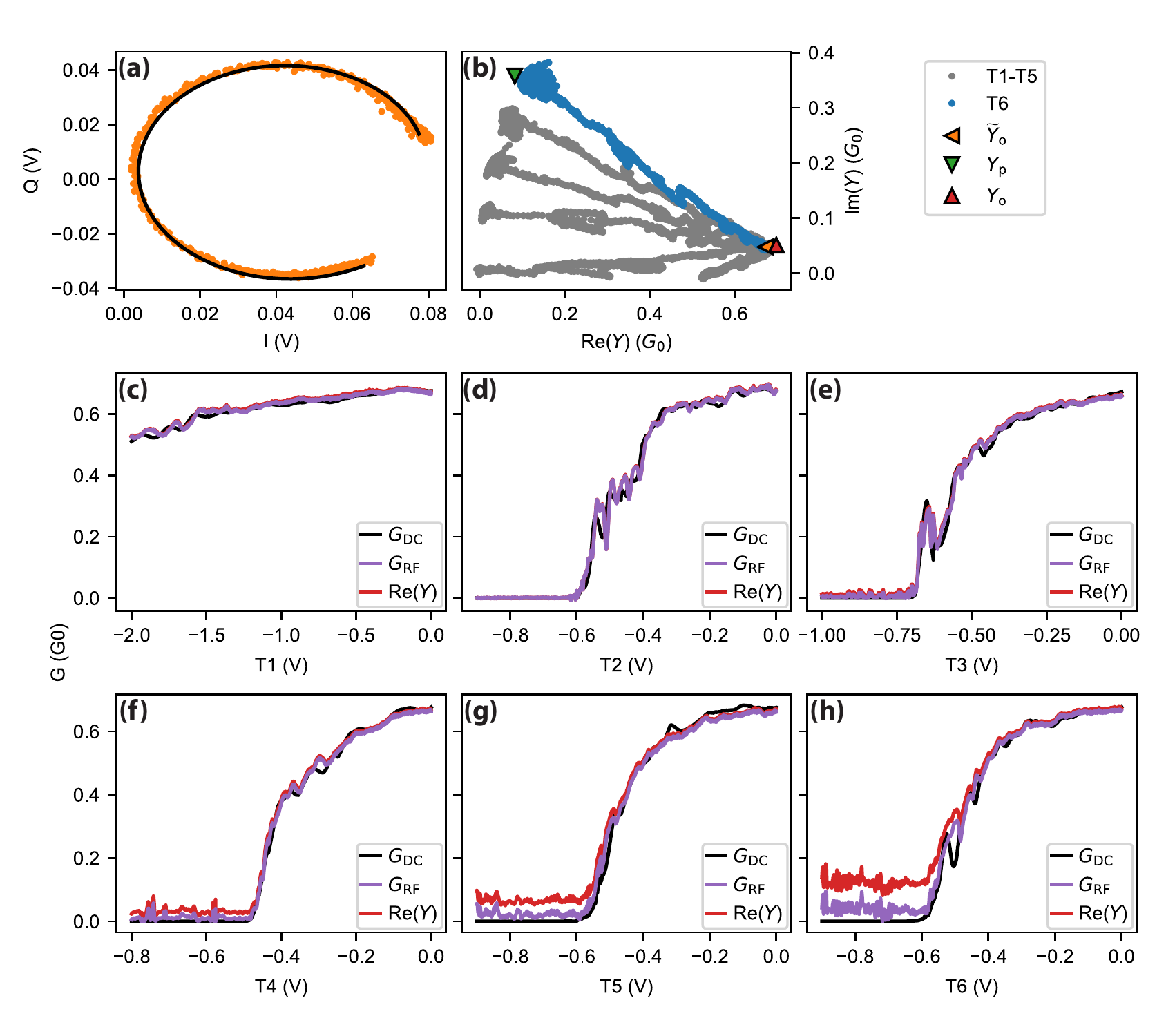}
    \caption{{\bf Supplemental data for pinch-off measurements.}
        {\bf a} Resonator response in the IQ plane together with a fit with Eq.~\eqref{eq:s21}. The phase delay of the line corresponding to $e^{\imath (\theta_0+\theta_1 \omega)}$ in Eq.~\eqref{eq:s21} is removed from both the data and the fit.
        {\bf b} Admittance data for all pinch-off measurements. The data obtained for T6 is highlighted in blue.
        {\bf c} to {\bf h} Conductance measured with DC techniques together with the conductance $G_\text{RF}$ extracted from the admittance in {\bf b}.
        To show the importance of correcting for the finite frequency effects in the nanowire, the real part of the admittance is also shown.
    }
    \label{fig:suppl_pinchoff}
\end{figure}

As described in the main paper, the determination of $\gamma \ell$ and $Z_\mathrm{nw}$ requires a measurement of the admittance in both the conducting and pinched-off regime.
Since there are many measurements of the admittance in both regimes, we here describe the procedure to fix $Y_\mathrm{p}$ and $Y_\mathrm{o}$.
To approach the open and pinched-off regimes as precisely as possible, the approach is to take the admittances that are furthest removed from the pinched-off and open regime respectively.
In practice, before we determine the admittance in the pinched-off regime of T6, $Y_\mathrm{p}$, we first select any admittance data point where all gates are open $\widetilde{Y}_\mathrm{o}$.
We then find $Y_\mathrm{p}$ as the point in the T6 pinchoff measurement that is furthest removed from $\widetilde{Y}_\mathrm{o}$.
Subsequently, we determine $Y_\mathrm{o}$ by finding the admittance furthest removed from $Y_\mathrm{p}$ in the aggregated data for all tunnel gates.
The aggregate data is shown in Fig.~\ref{fig:suppl_pinchoff}b, with the datasets from the T6 pinchoff measurement highlighted in blue.
The obtained points $\widetilde{Y}_\mathrm{o}$, $Y_\mathrm{p}$ and $Y_\mathrm{o}$ are also identified in the figure.
We obtain $\gamma \ell = 0.6+0.3\imath$ and $Z_\mathrm{nw}= (16.7+3.6\imath)\,\SI{}{\kilo\ohm}$.

\section{Comparison of $G_\text{DC}$ and $G_\text{RF}$ for all pinchoff curves}
Here, we provide a more detailed overview of all admittance data obtained and used for the inset in Fig.~2e.
Similar to Fig.~2e, we plot a comparison between $G_\text{DC}$ and $G_\text{RF}$ for all tunnel gates in Fig.~\ref{fig:suppl_pinchoff}c-h.
Additionally, we therein show the real part of the admittance, equivalent by definition to the finite frequency conductance.
The further the tunnel gate is from the source of the nanowire, the more important the correction for finite frequency effects is to obtain the correct $G_\text{RF}$.
This is expected since a larger portion of the shunting capacitance is available as an alternative path to ground which becomes more dominant the closer the nanowire is to pinch-off.

\section{Relation between SNR and amplifier noise temperature}
To calculate the relation between SNR and the equivalent noise temperature of the amplifier, we assume that the noise level of the input signal is negligible.
The equivalent noise temperature $T_\mathrm{N}$, is defined as \cite{Pozar2011}
\begin{equation}
T_\mathrm{N} = \frac{N_\mathrm{in}}{k_\mathrm{B} B},
\end{equation}
where $B$ denotes the measurement bandwidth and $N_\mathrm{in}$ the equivalent noise input power to the amplifier.
Since the integration time, $t_\mathrm{m}$ is longer than any other timescale in the system, the bandwidth is given by $B=1/t_\mathrm{m}$.

We calculate the voltage fluctuations corresponding to this noise power as
$v = \sqrt{N_\mathrm{in} Z},$
where $Z$ is the characteristic impedance of the feedline.
Using $G_\mathrm{sys}$ to denote the gain of the amplification in the system, the SNR is defined as the ratio between signal $\Delta = G_\mathrm{sys} \Delta_\mathrm{f}$ and the noise $2 G_\mathrm{sys} v$.
As such, we find the following equation for the SNR 
\begin{equation}
\mathrm{SNR} = \frac{G_\mathrm{sys} \Delta_\mathrm{f}}{2 G_\mathrm{sys} v} = \frac{\Delta_\mathrm{f}\sqrt{t_\mathrm{m}}}{2 \sqrt{Z k_\mathrm{B} T_\mathrm{N}}},
\end{equation}
assuming the SNR is limited by the noise introduced by the finite noise temperature of the amplifiers in the system.